\documentclass[a4paper,twoside]{article}
\usepackage{graphicx}
\usepackage{amsmath}

\title{A two species trap for chromium and rubidium atoms}
\author{Sven Hensler, Axel Griesmaier, J\"org Werner, Axel G\"orlitz and Tilman Pfau\\5. Physikalisches Institut, Universit\"at Stuttgart, 70550 Stuttgart (Germany)}
\begin{document}

\maketitle
\begin{abstract}
{We realize a combined trap for bosonic chromium ($^{52}$Cr) and
rubidium ($^{87}$Rb) atoms. First experiments focus on exploring a
suitable loading scheme for the combined trap and on studies of
new trap loss mechanisms originating from simultaneous trapping of
two species. By comparing the trap loss from the $^{87}$Rb
magneto-optical trap (MOT) in absence and presence of magnetically
trapped ground state $^{52}$Cr atoms we determine the scattering
cross section of $\sigma_{inelRbCr}=5.0\pm4.0\cdot10^{-18}$\,m$^2$
for light induced inelastic collisions between the two species.
Studying the trap loss from the Rb magneto-optical trap induced by
the Cr cooling-laser light, the photoionization cross section of
the excited $5$P$_{3/2}$ state at an ionizing wavelength of
426\,nm is measured to be
$\sigma_{p}=1.1\pm0.3\cdot10^{-21}$\,m$^2$.}
\end{abstract}
\section{Introduction}
The dominant interaction in atomic Bose-Einstein condensates (BEC)
realized so far, is the contact interaction. In the ultracold
regime, this interaction is isotropic and short range. Recently,
increasing theoretical interest has focused on the dipole-dipole
interaction \cite{Baranov2002,Goral2000,San00}. This interaction
is long range and anisotropic and thus would greatly enrich the
physics of degenerate quantum gases. Additionally, tuning of this
interaction from attraction to repulsion is possible by applying
time dependent external fields \cite{Gio02}. Together with the use
of a Feshbach resonance to vary the s-wave scattering length this
would allow control of the scattering properties of an ultracold
sample over a wide range of values and characteristics.

A promising candidate for the observation of the influence of the
dipole-dipole interaction on the dynamics of a BEC is atomic
chromium \cite{Schmidt2003}. Due to its comparably large magnetic
dipole moment of 6$\mu_B$, where $\mu_B$ is the Bohr magneton, the
dipole-dipole interaction is of the same order of magnitude as the
contact interaction. However, many of the effects proposed for
dipolar gases \cite{Pu01,demille,Baranov2002b} require a much
stronger dipole-dipole interaction. Magnetic dipole moments or
electric dipole moments aligned in external fields that lead to
such strong dipolar interaction can be found in heteronuclear
molecules. The strongly paramagnetic Cr-Rb-molecule, is thus
expected to have a large electric dipole moment beside its
magnetic moment. In addition, the generation of degenerate gases
of ultracold Cr-Rb-molecules formed by two ultracold gases of Cr
and Rb which are each produced by laser cooling and subsequent
sympathetic cooling \cite{Modugno2001} seems to be feasible.

The paper is organized as follows. After a short description of
our experimental setup, we present our findings on photoionization
of the excited Rb$^*$. Measurements concerning interspecies
collisions can be found in the subsequent section followed by our
conclusions.

\section{The combined magneto-optical trap for Cr and Rb}
As a first step, on the way to the generation of ultracold,
heteronuclear molecules, we have realized a two species
magneto-optical trap for chromium and rubidium atoms. For our
measurement we prepare both MOTs in the quadrupol field of two
coils oriented in anti-Helmholtz configuration. Each trap is
formed by three orthogonal and retro-reflected trapping laser
beams, with the typical $\sigma^+-\sigma^-$-polarization. The
trapping beams of the two traps are tilted by a small angle with
respect to each other to be able to set up separate optics for
both traps. We load about $N_{Cr}=4\cdot10^6$ bosonic
$^{52}$Cr-atoms from a high temperature effusion cell via a
Zeeman-slower into the Cr-MOT. The 426\,nm cooling light is
generated by a frequency doubled Ti:Sa-laser. The Rb-MOT with a
steady state atom number of about $N_{Rb}=3\cdot10^6$ Rb-atoms is
loaded from the Rb-background gas provided by a continuously
operated Rb-getter source. The stabilized Rb-cooling and repumping
lasers are provided via a fiber from a different experimental
setup. Both frequencies separated by 6.8\,GHz are amplified by a
single laser diode using injection locking technique. When both
traps are operated at the same time, the steady state number of Rb
atoms $N_{Rb}$ drops by a factor of 5 due to photoionization of
the excited Rb-cooling state.

\section{Photoionization of magneto-optically trapped rubidium atoms}
Photoionization of a neutral atom is a transition from a bound
state $|i\rangle$ with internal energy $E_i$ to a continuum state
$|f\rangle$. Such a transition becomes possible if the energy
${E_p=\hbar\omega_p}$ of an incident photon exceeds the
ionization-energy ${E_{ion}=\hbar\omega_{ion}}$ of the bound state
$|i\rangle$. The transition rate ${\Gamma_{i{\rightarrow}f}}$ is
given by Fermi's Golden Rule:
\begin{equation}
\label{gammaif}
\Gamma_{i{\rightarrow}f}=\frac{2\pi}{\hbar}\left|\left<f\left|\hat{H}_{ia}\right|i\right>\right|^2\rho(E_{i}+E_p),
\end{equation}
where $\left<f\left|\hat{H}_{ia}\right|i\right>$ is the transition
matrix element of the interaction Hamiltonian $\hat{H}_{ia}$ and
$\rho(E_{i}+E_p)$ is the density of states in the continuum at the
final energy $E_{i}+E_p$. The photoionization rate can also be
expressed using the photoionization cross-section
$\sigma(\omega_p)$ if we define the photon-flux
$\Phi=\frac{I_p}{\hbar\omega_p}$ using the intensity $I_p$ of the
ionizing light with a frequency of $\omega_p$:
\begin{equation}\label{blueint}
\Gamma_{i{\rightarrow}f}(\omega_p,I_p)=\sigma_p(\omega_p)\Phi
\end{equation}
\begin{figure}[t]
\begin{minipage}[t]{0.48\linewidth}
    \centering
    \includegraphics[width=\textwidth]{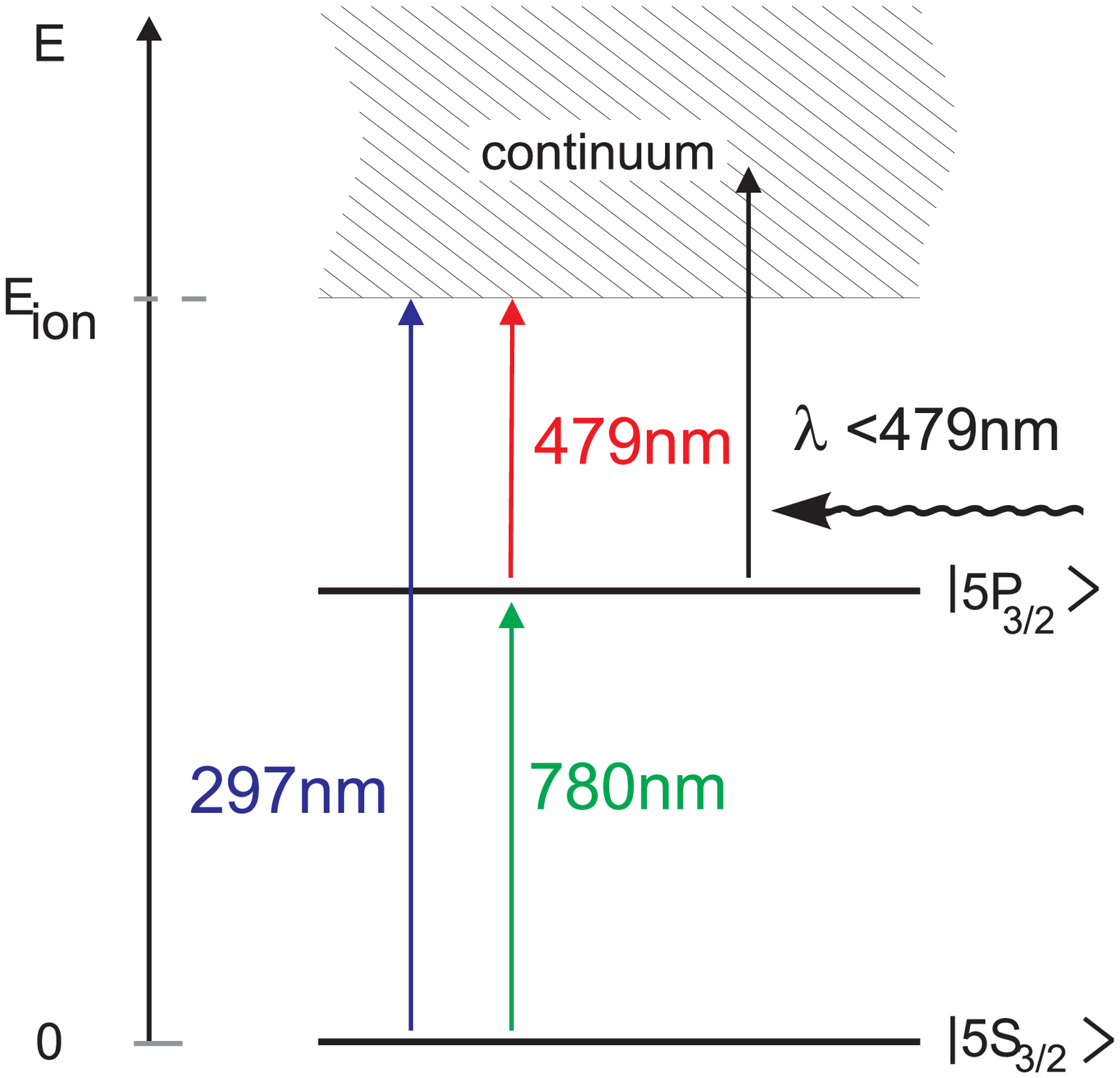}\\
    \caption{Energy diagram of the photoionization of Rubidium:
    The energy of the incident photon is sufficient to ionize the excited
    $5$P$_{3/2}$ state. The $5$S$_{1/2}$ ground state is unaffected. Above $E_{ion}$ the possible
    final states form a continuum of states.}\label{pischeme}
  \end{minipage}%
  \hfill
  \begin{minipage}[t]{0.48\linewidth}
    \centering
    \includegraphics[width=\textwidth]{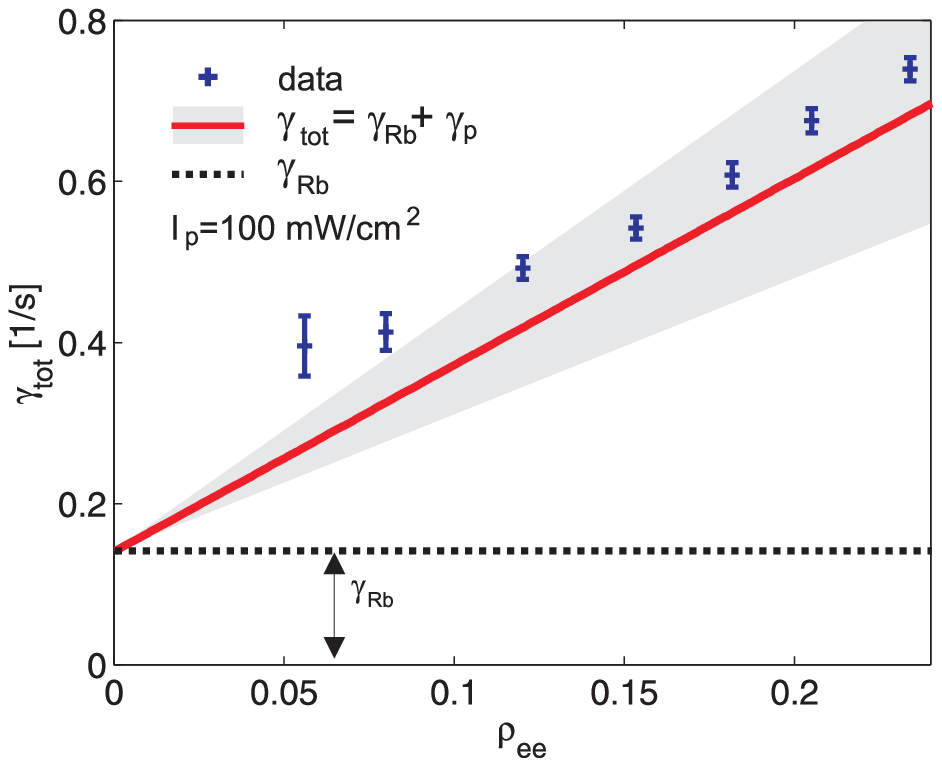}\\
    \caption{Dependence of the one body loss rate of the $Rb$ trap on the population ${\rho_{ee}}$ of the excited 5P$_{3/2}$ state at an intensity of the
    ionizing laser of 100\,mW/cm$^2$. Note, the straight line is not a fit to the data, it is a calculated line using
    the mean value $\sigma_{p}=1.1\pm0.3\cdot10^{-21}$\,m$^2$ and the experimental
    parameters, where the error of $\sigma_{p}$ is indicated by a grey area.
    The dotted line represents the mean of the measured pure background loss rates $\gamma_{Rb}$.}\label{gammatotlin}
  \end{minipage}
\end{figure}

For $^{87}$Rb the ionization threshold is at an energy level of
4.2\,eV (297\,nm) above the $5$S$_{1/2}$ ground state. With an
energy difference of 1.6\,eV (780\,nm) between the ground state
and the $5$P$_{3/2}$ excited state, the ionization energy of the
excited state is 2.6\,eV (479\,nm). Thus, if we apply the Cr
cooling laser (426\,nm) in the Rb-MOT, transitions of excited
Rb$^*$ atoms to continuum states become possible. This situation
is displayed in figure \ref{pischeme}. During the process, energy
and momentum are conserved and the excess energy is distributed
among the electron and the ion leading to a velocity of the
electron of $v_e\approx3.4\cdot10^5$\,m/s. Therefore,
recombination within the trap volume is very unlikely. Ions and
electrons are not supported by the trap. This additional loss
contributes to the total one-body loss rate, which reads
$\gamma_{tot}=\gamma_{Rb}+\gamma_{p}$, where $\gamma_{p}$ is the
loss rate induced by the ionization and $\gamma_{Rb}$ represents
all other one body losses which are mainly caused by background
gas collisions. As the ionization occurs in the excited state, the
ionization rate $\Gamma_{i{\rightarrow}f}$ has to be multiplied
with the population probability $\rho_{ee}$ of the excited state:
\begin{equation}
\label{gammaeq} \gamma_{p}=\Gamma_{i{\rightarrow}f} \rho_{ee}
\qquad, \mathrm{where}
\end{equation}
\begin{equation}
\label{rhoeeeq} \rho_{ee}=\frac{s}{2({s+1})},\quad s=\frac{\langle
C\rangle^2 I/I_s}{1+(2\delta/\Gamma)^2}.
\end{equation}
Here $s$ is the saturation parameter expressed by the average
Clebsch-Gordan coefficient $\langle C\rangle^2=7/15$, $\Gamma$ the
natural linewidth and $I_s=1.6\,$mW/cm$^2$ the saturation
intensity of the Rb trapping transition in a light field with
total intensity $I$ and detuning $\delta$.

Neglecting two and three-body losses, which is a good
approximation for the low  densities we observe in our trap, the
time evolution of the atom number $N_{Rb}(t)$ in the rubidium trap
can be described by the following well known rate equation:
\begin{equation}
\frac{dN_{Rb}}{dt}=L-\gamma_{tot} N_{Rb},
\end{equation}
where $L$ is the loading rate. We record loading curves of the
Rb-MOT with a photodiode at different intensities of the Rb-MOT
beams ranging from 20\,mW/cm$^2$ to 160\,mW/cm$^2$ and intensities
of the Cr trapping light from 0\,mW/cm$^2$ to 600\,mW/cm$^2$ and
determine the total loss rate $\gamma_{tot}$ under each of these
conditions. In figure \ref{gammatotlin} the measured loss rate
$\gamma_{tot}$ is plotted over $\rho_{ee}$ at a total intensity of
the ionizing laser of ${I_p=}$100\,mW/cm$^2$. Subtracting the
background loss rate $\gamma_{rb}$ obtained from loading curves
with no ionizing light present from the total loss rate, we
extract the ionization rate
${\gamma_{p}(I_{Rb})=\gamma_{tot}(I_{Rb})-\gamma_{Rb}}$. The
ionization rate shows a linear increase with the excited state
population and vanishes for $\rho_{ee}\rightarrow0$. We therefore
attribute the increase of the one-body loss rate when the 426\,nm
Cr-trapping light is switched on to photoionization of excited
Rb$^*$ atoms.

The population of the excited state is calculated using Eq.
(\ref{rhoeeeq}) and the intensities of the beams which are
measured outside the chamber and corrected taking into account the
transmission of the windows and the retroreflected beams. The
Rb-MOT laser is tuned to maximum intensity of the fluorescence of
the MOT, suggesting a detuning\footnote{According to several
publications (see e.g. \cite{Rapol2001}) maximum fluorescence is
observed at a detuning of $\delta\approx2.25\Gamma$ and we
estimated an accuracy of $\pm0.25\Gamma$ of this value.} of
$\delta\approx2.25\pm0.25\Gamma$. We estimate a total systematic
error caused by the uncertainty in the detuning and spatial
inhomogeneity of the laser beams of 20$\%$. For each intensity of
the Rb MOT beams a least square linear fit of Eq. (\ref{blueint})
to the measured rate constants $\Gamma_{i\rightarrow
f}=\gamma_{p}/\rho_{ee}$ yields $\sigma_{p}$. The mean value of
these photoionization cross sections of the $5P$ state of
$^{87}$Rb at a wavelength of 426\,nm is:
\begin{equation}
\sigma_{p}(426nm)=1.1\pm0.3\cdot10^{-21}\mathrm{m}^2
\end{equation}
This value is in good agreement with previously published values
\cite{Din92,Fuso2000} at comparable wavelengths and with
theoretical predictions \cite{Aym83}.

\section{Inelastic interspecies collisions}
In the second part of this article, we investigate the inelastic
trap losses in combined Cr-Rb traps. During an interspecies
collision in the presence of cooling light, each atom (Cr and Rb)
undergoes several transitions from the ground to the excited
state. If the atoms approach each other they experience a
molecular potential. For ground state atoms the long-range part of
this potential is dominated by the attractive van-der-Waals
potential $V_{gg}(r)=C_6/r^6$, where r is the internuclear
distance. Based on a two level model Schl\"oder et al.
\cite{Schloeder1999} pointed out, that this part of the molecular
potential does not change its dependence on r if one of the atoms
is in the excited state $V_{eg}(r)=C^*_6/r^6$. The $C^*_6$
coefficient leads to an attractive (repulsive) interaction, if the
excited atom has the smaller (larger) transition frequency of both
the atoms. Since $C^*_6$ is always larger than $C_6$, the excited
state potential is steeper. This has two consequences for Cr-Rb
collisions:

First, inelastic collisions involving the excited state of $Cr^*$
are prevented, because the steep, repulsive excited state
potential hinders the two species to get very close to each other.
Second, in contrast to homonuclear collisions, where the potential
$V_{eg}$ is proportional to $r^{-3}$, in the heteronuclear case,
the colliding atoms decouple from the light field at smaller
internuclear distances. This leads to a higher survival
probability \cite{Gallagher1989,Weiner1999}, which means that the
two atoms can approach each other in the Rb$^*$-Cr-potential to a
internuclear distance where fine structure changing collisions
(FC) and radiative escape (RE) lead to trap losses.

Since loss due to photoionization of excited Rb-atoms triggered by
the Cr-trapping light and loss due to Cr-light induced two-body
collisions in the Cr-MOT \cite{Bradley2000} are dominant in both
MOTs, it was not possible to observe trap loss due to interspecies
collisions during the simultaneous operation of these traps.
However, because of the 6 times larger magnetic moment of Cr we
are able to prepare a magnetically trapped (MT) cloud of Cr-atoms
and study the interaction with magneto-optically trapped Rb-atoms
in the absence of the Cr-cooling light. As explained above, in a
combined MOT for Cr and Rb we do not expect excited Cr atoms to
contribute to the inelastic loss coefficient, therefore, the
measured $\beta$-coefficient for inelastic collisions in the
overlapped MT for Cr and MOT for Rb should be very similar to a
coefficient measured while both MOTs are operated simultaneously.

For this measurement, we magnetically trap about $5\cdot10^7$
$^{52}$Cr-atoms at $100\,\mu$K in the $^7$S$_3$ ground state using
the continuous loading scheme presented by Stuhler et al.
\cite{Stu01}. Here, the MT potential is created by the same coil
configuration as described in the previous section. During the
measurements the field gradient in the direction of the coil axis
is $25$\,G/cm. Assuming a cut off parameter of $\eta=10$ this
results in a calculated magnetic trap depth of
$k_B\cdot530$\,$\mu$K for Cr-atoms in the extreme Zeeman substate
\cite{Luiten1996}. We then apply the Rb cooling and repumping
light for a certain interaction time $t$ to load the Rb-MOT. The
temperature of the Rb cloud is about $320\,\mu$K. The light forces
in the Rb-MOT result in a much deeper trapping potential of
approximately $k_B\cdot8$\,K. After the interaction time $t$ the
fluorescence of either the Rb-cloud or of the resonantly
illuminated Cr-atoms is imaged onto a calibrated CCD-camera while
no near resonant light is applied to the other species. The number
of atoms of the imaged species is calculated from the pixel count
of the image. We perform a series of such measurements in which we
record the evolution of the number of Cr-atoms $N_{Cr}(t)$ and
Rb-atoms $N_{Rb}(t)$. The temperature and the magnetic moment of
the magnetically trapped Cr-Atoms are obtained from a 2D-fit to
the atomic distribution in a quadrupole field under the influence
of the gravity. The temperature of the Rb-atoms is deduced from
the ballistic expansion of the cloud.

\begin{figure}[t]
\begin{minipage}[t]{0.48\linewidth}
    \centering \includegraphics[width=\textwidth]{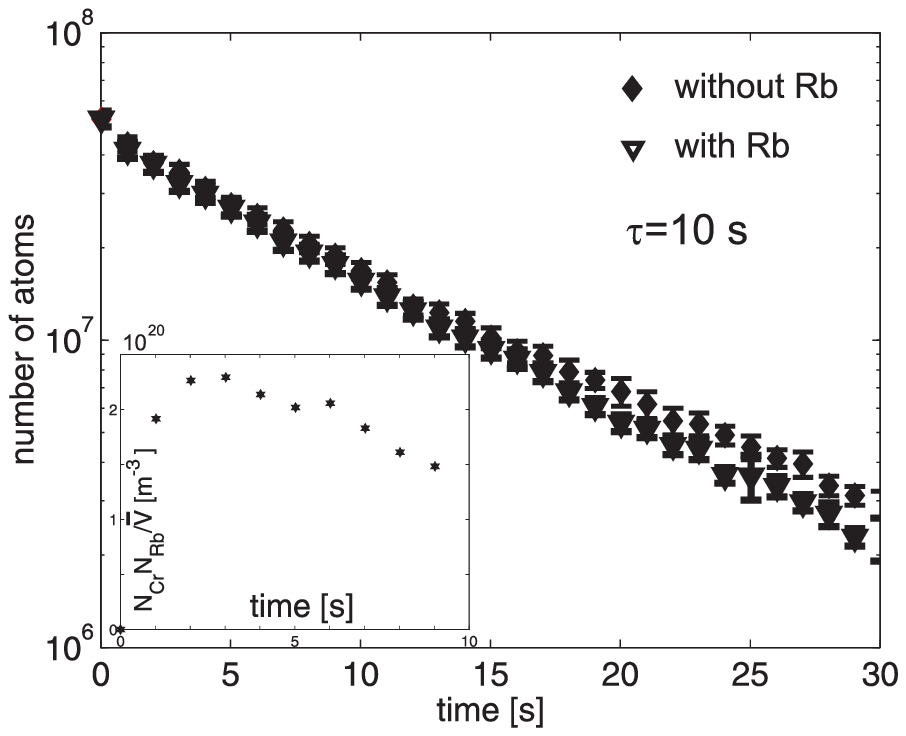}\\
  \caption{Decay of the number of magnetically trapped Cr-atoms with and without the influence of the Rb-MOT.
  The inset indicates the constants of the factor $N_{Cr}N_{Rb}/\overline{V}$.}\label{fig:zerfall_cr}
  \end{minipage}%
  \hfill
  \begin{minipage}[t]{0.48\linewidth}
    \centering \includegraphics[width=\textwidth]{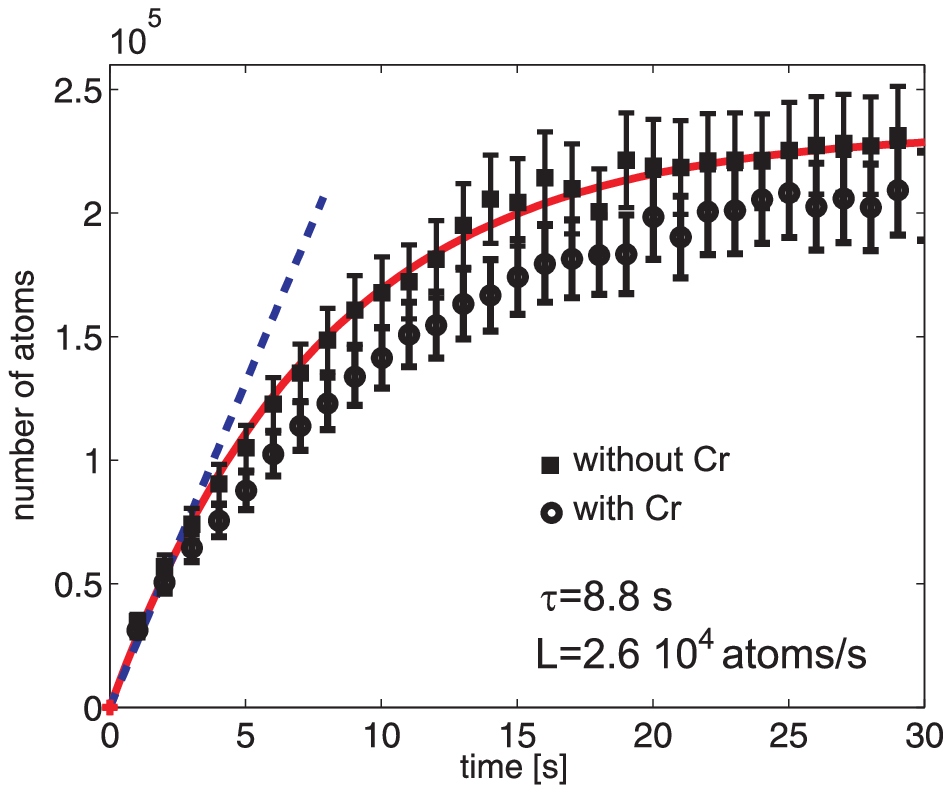}\\
  \caption{Loading curve of the Rb-MOT in the presence and absence of magnetically trapped Cr-atoms. The solid line indicates
  a fit of the one-body loading equation to the measurement without magnetically trapped Cr-atoms.
  The dotted line represents the initial slope of this fit.}\label{fig:laden_rb}
  \end{minipage}
\end{figure}

Figure \ref{fig:zerfall_cr} illustrates the decay of the number of
magnetically trapped Cr-atoms during the first 30\,s with and
without trapped Rb-atoms being present. Neglecting two-body loss,
a fit of a one-body decay without trapped Rb-atoms present yields
a lifetime of 10\,s. In the presence of the Rb-MOT the atom number
is reduced by $8\cdot10^5$\,atoms (25\%) after 30\,s.

The loading process of the Rb-MOT is documented in figure
\ref{fig:laden_rb}. The measurement is performed with and without
magnetically trapped Cr-atoms. Neglecting two- and three-body
processes, we obtain a loading rate of
$L_{Rb}=2.6\cdot10^4$\,atoms/s and a lifetime of $9$\,s from a fit
of the one-body loading equation to the data without trapped
Cr-atoms. This yields a steady-state atom number of
$2.3\cdot10^5$\,atoms. The additional loss channel introduced by
magnetically trapped Cr-atoms can be clearly observed.

The decay of the number of magnetically trapped Cr-atoms
$N_{Cr}(t)$ and the increase of magneto-optically trapped Rb-atoms
$N_{Rb}(t)$ during loading in the presence of the other species is
governed by the following coupled rate equations:
\begin{eqnarray}
    \label{eq:ratengleichungenCr}\frac{dN_{Cr}}{dt}&=&-\gamma_{Cr} N_{Cr}-\beta_{CrRb}\int d^3r\,n_{Cr} n_{Rb} \\
    \label{eq:ratengleichungenRb}\frac{dN_{Rb}}{dt}&=&L_{Rb}-\gamma_{Rb} N_{Rb}- \beta_{RbCr}\int d^3r\,n_{Cr} n_{Rb}
\end{eqnarray}
Because of the different loss channels of the two types of traps,
$\beta_{CrRb}$ and $\beta_{RbCr}$ are not expected to be equal.
The integral over both density distributions in the interspecies
collision term can be approximated by the following expression
\cite{Stuhler2001b}:
\begin{eqnarray}\label{eq:ueberlappvolumen}
    \int d^3r\, n_{Cr} n_{Rb}&=& \frac{N_{Cr} N_{Rb}}{\overline{V}},\\
    \overline{V}&=&V_{MT}/\varsigma, \\
    \varsigma&=&e^{\frac{\overline{\sigma}^2}{2
    z^2}}\left(\frac{\overline{\sigma}^2}{z^2}+1\right)\left(1-\mathrm{Erf}
    \left[\frac{\overline{\sigma}}{\sqrt{2}z}\right]\right)-
    \sqrt{\frac{2}{\pi}}\frac{\overline{\sigma}}{z},
\end{eqnarray}
where the effective collision volume $\overline{V}$ varies between
the Cr-MT volume $V_{MT}=8\pi z^3$ for $\overline{\sigma}\ll z$
and the volume of the Rb-MOT $V_{MOT}$ for $\overline{\sigma}\gg
z$. The magnetic trap and the MOT are regarded to be isotropic
with a $1/e$-length $z$ for the Cr cloud in the MT in direction of
the coils axes and a mean $1/\sqrt{e}$-size $\overline{\sigma}$
for the Rb-cloud in the MOT, respectively.

In order to deduce the loss coefficient $\beta_{RbCr}$ from the Rb
loading curve, Eq. (\ref{eq:ratengleichungenRb}) is solved for the
first seconds by assuming a constant factor
$N_{Cr}N_{Rb}/\overline{V}$ which is well reproduced by our data
due to the inverse evolution of the atom numbers (see inset figure
\ref{fig:zerfall_cr}). From the initial slope $\alpha$, the loss
coefficient $\beta_{RbCr}$ can be calculated:
\begin{equation}\label{eq:slope}
    \alpha=L_{Rb}-\beta_{RbCr}\frac{N_{Cr}N_{Rb}}{\overline{V}}.
\end{equation}
Using $L_{Rb}$ from the measurements in which no Cr-atom was
trapped we obtain a loss coefficient of
$\beta_{RbCr}=1.4\pm1.1\cdot10^{-17}$\,m$^3$/s with a population
probability of the excited Rb state of about 25\%. This yields an
inelastic cross section of $\sigma_{inel,RbCr}=\beta_{RbCr}
\overline{v}=5.0\pm4.0\cdot 10^{-18}$\,m$^2$, where we have used
the mean velocity $\overline{v}=\sqrt{\frac{8
k_B}{\pi}\left(\frac{T_{Cr}}{m_{Cr}}+\frac{T_{Rb}}{m_{Rb}}\right)}$.
Due to uncertainties in $N_{Cr}, N_{Rb}$ and $\overline{V}$ we
estimate a relative systematic error of 80\%.

From the difference in the number of Cr atoms in the presence and
absence of the trapped Rb cloud, we extract an upper and lower
limit for the loss coefficient taking the maximum and minimum
value of the factor $N_{Cr}N_{Rb}/\overline{V}$ and assuming this
factor to be constant over the first 30\,s:
$4.7\cdot10^{-16}\,$m$^3$/s $< \beta_{CrRb}<
5.5\cdot10^{-15}\,$m$^3$/s. For this approximation we used the
data points between 20\,s and 30\,s, where the data are well
separated. The systematic relative error of these limits is again
about 80\%. The loss coefficients $\beta_{RbCr}$ and
$\beta_{CrRb}$ we obtain, thus, differ by one order of magnitude.
We attribute this to the large difference of the trap depths of
the dissipative Rb-MOT and the conservative Cr-MT. Losses arise if
the energy gained by inelastic collisions between unpolarized Rb
and polarized Cr atoms is sufficiently high to eject an atom from
its trap. Due to energy and momentum conservation 63\% of the
released energy is transferred to the Cr atoms. While in the
Rb-MOT only fine structure changing (FC) and radiative escape (RE)
interspecies collisions lead to additional trap loss, in the
shallow Cr-MT FC, RE and hyperfine changing collision in the Rb
atom, interspecies dipolar relaxation and depolarizing collisions
which end in un-trapped states reduce the Cr atom number.

\begin{figure}
  \centering
  \includegraphics[width=0.5\textwidth]{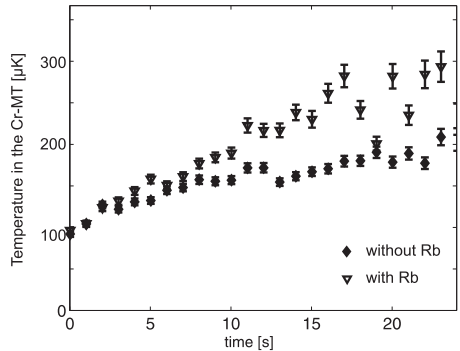}\\
  \caption{Temperature increase of the Cr-cloud in the MT. Depicted is the
  temperature evolution of the cloud in presence and absence of the Rb-MOT}\label{fig:tempincr}
\end{figure}

The atom loss in the Cr-MT in the presence of the Rb-MOT is
accompanied by a temperature increase ($\sim 8\,\mu$K) of the Cr
cloud. Figure \ref{fig:tempincr} depicts the evolution of the
temperature with and without the presence of the other species.
Without Rb atoms present, this heating rate is reduced to
$4\,\mu$K/s and is mainly caused by anti-evaporation due to
Majorana losses in the Cr-trap. We exclude thermalization with the
Rb-MOT as the cause of the additional heating, because the same
heating rate of the Cr cloud was measured in Cr traps prepared at
temperatures lower and higher than the temperature of the Rb-MOT.
Since the cloud is not collisionally dense, the atoms are expelled
from the trap without depositing any energy in the cloud after
most inelastic processes. Therefore, we attribute this increase to
anti-evaporation caused by the mentioned loss mechanism within an
inhomogeneous gas of Rb atoms and to collisions which lead to
depolarization of the magnetic sublevels. In latter collisions,
the energy difference between the substates of Cr and Rb (Rb$^*$)
which is $3/2\mu_B B$ ($4/3\mu_B B$) in the small magnetic field
near the trap center, is released and heats up the cloud.

\section{Conclusions}
The first results on a combined Cr-Rb trap which are presented in
this paper show that the operation of both MOTs is limited by the
photoionization of the excited state of the Rb-atoms which occurs
with a cross section of $\sigma_{p}=1.1\pm 0.3\cdot10^{-21}m^2$.
Due to other dominant loss mechanisms interspecies collisions
could not be directly observed in the combined MOTs. By
overlapping the Rb-MOT and the Cr-MT in space and time, we
measured light induced interspecies collisions with a cross
section of $\sigma_{inel,RbCr}=5.0\pm 4.0\cdot10^{-18}$m$^2$, in
the Rb-MOT. Since collisions involving the excited state of Cr are
prevented in a combined MOT of Rb and Cr atoms, a very similar
cross section is expected for light induced collisions in a
combined MOT. Due to the contribution of additional loss channels,
the loss rate of Cr-atoms in the very shallow MT is more
pronounced. Simultaneous operation of both MOTs could be improved
by alternating cooling laser pulses for the two species which
would suppress photoionization. If a Rb ground state trap is
prepared before the Cr-MOT is loaded, light-induced interspecies
collisions could be prevented.

The discussed measurements have already indicated the richness of
the interaction in a combined system of trapped Cr and Rb atoms.
An improvement of our Rb-source should allow us to load a
significant number of Rb-atoms into a Rb-MT and study ground state
collisions. These measurements will allow us to extract the
elastic and inelastic interspecies ground state cross sections
which are important for sympathetic cooling. Here, studies of
inelastic processes resulting from the interaction of two species
with very different internal structures are of theoretical
interest to gain a deeper understanding of these relaxation
processes.

\section{Acknowledgment}
We thank Axel Grabowski for providing us with stabilized Rb
trapping light, K. Rz\c{a}\.{z}ewski and J. Stuhler for fruitful
discussions. This work was funded by the DFG SPP 1116 and the RTN
network "Cold Quantum Gases" under the contract No.
HPRN-CT-2000-00125.

%% bibliography

\end{document}